\documentclass[12pt]{article} 
\usepackage{amsmath,amssymb} 
\usepackage{graphicx} 
\usepackage{color} 
\usepackage{hyperref} 
\hypersetup{bookmarksopen=true, bookmarksnumbered=true, pdfstartview={FitH}, pdfborder={0 0 0}, colorlinks=true, linkcolor=red, linktocpage=true, citecolor=blue, urlcolor=cyan, unicode=true, pdftitle={Massive Scalar Hypermultiplet in Projective Hyperspace}, pdfauthor={DJain}, pdfsubject={Projective Hyperspace}, pdfkeywords={Hypermultiplet, }{Hyperspace, }{Central Charge}}
\usepackage{geometry} 
\usepackage{fancyhdr} 
\usepackage[nottoc,notlof,notlot]{tocbibind} 
\usepackage[titles]{tocloft} 

\usepackage{parskip} 
\definecolor{title}{rgb}{0.7,0.9,0.1}
\definecolor{abst}{rgb}{0.366,0.366,0.266}
\definecolor{sect}{rgb}{1.0,0.0,0.0}
\definecolor{ssect}{rgb}{0.5,0.5,0.0}
\definecolor{sssect}{rgb}{0.3,0.3,0.3}
\definecolor{appsect}{rgb}{0.0,1.0,0.0}
\definecolor{ref}{rgb}{0.0,0.0,1.0}

\newcommand\sect[1] {{\color{sect}\section{#1}}}
\newcommand\subsect[1] {{\color{ssect}\subsection{#1}}}

\newcommand\references[1] {\color{ref} \begin{thebibliography}{[99]} \color{black} #1 \end{thebibliography}}
\numberwithin{equation}{section} 
\newcommand\bc {\begin{center}}
\newcommand\ec {\end{center}}
\newcommand\be {\begin{equation}}
\newcommand\ee {\end{equation}}
\newcommand\bfig {\begin{figure}}
\newcommand\efig {\end{figure}}
\newcommand\bpm {\begin{pmatrix}}
\newcommand\epm {\end{pmatrix}}
\renewcommand\({\left(}
\renewcommand\){\right)}
\renewcommand{\exp}{e^}
\newcommand{\I}{\dot{\iota}}
\newcommand\A {{\cal A}}
\newcommand\D {{\cal D}}
\renewcommand\S {{\cal S}}
\renewcommand\' {^{\prime}}

\newcommand\arXivid[1] {\href{http://arxiv.org/abs/#1}{\tt ar$\chi$iv:#1}} 
\newcommand\cmp[3] {{\it Commun.\ Phys.\ Math.\ }{\bf #1} (#2) #3} 
\newcommand\cqg[3] {{\it Class.\ Quant.\ Grav.\ }{\bf #1} (#2) #3}
\newcommand\jhep[3]{{\it JHEP\ }{\bf #1} (#2) #3}
\newcommand\npb[3] {{\it Nucl.\ Phys.\ }{\bf B #1} (#2) #3}
\newcommand\pr[4] {{\it Phys.\ Rev.\ }{\bf #1 #2} (#3) #4} 

\begin{document}
\pagenumbering{alph}
\title{\color{title}\Huge A Note on Massive Scalar Hypermultiplet in Projective Hyperspace}
\author{Dharmesh Jain\footnote{\href{mailto:djain@insti.physics.sunysb.edu}{djain@insti.physics.sunysb.edu}}\, , Warren Siegel\footnote{\href{mailto:siegel@insti.physics.sunysb.edu}{siegel@insti.physics.sunysb.edu}\vskip 0pt \hskip 8pt \href{http://insti.physics.sunysb.edu/\~siegel/plan.html}{http://insti.physics.sunysb.edu/$\sim$siegel/plan.html}}\bigskip\\ \emph{C. N. Yang Institute for Theoretical Physics}\\ \emph{State University of New York, Stony Brook, NY 11790-3840}}
\date{} 
\maketitle
\thispagestyle{fancy}
\rhead{YITP-SB-11-19} 
\lhead{\today}
\begin{abstract}
\normalsize We analyze the massive 4D scalar multiplet in `reformulated' projective N=2 superspace (hyperspace) from both 4D and 6D perspectives.
\end{abstract}

\tableofcontents

\newpage
\cfoot{\thepage}\rhead{}\lhead{}
\pagenumbering{arabic}
\sect{Introduction}
Introducing central charges in superalgebras leads to the possibility of having massive multiplets as `short' as the massless ones. The central charges in 4D, N=2 superspace have been dealt directly in both Projective\cite{ULMR,FGrey} \& Harmonic\cite{GIKOS,IBSK} hyperspaces.

The projective hyperspace has recently been formulated in coset space language in \cite{WS-AdS}, which has been used to simplify derivation of earlier results and perform new calculations involving massless scalar and vector hypermultiplets in \cite{DJWS}. For the sake of completeness, in this note we extend such an analysis to the massive case.

In the next section, we review the projective hyperspace with central charges. Then we discuss the massive scalar hypermultiplet in detail from the 4D perspective. Next, we show that the dimensional reduction of a massless hypermultiplet from 6D reproduces all the 4D results rather trivially. Finally, we present a simple 1-hoop calculation using Feynman rules similar to the massless case.

\sect{Projective Hyperspace with Central Charges\label{sec2}}
We use the conventions of \cite{DJWS} for the superspace coordinates and derivatives. The centrally extended algebra of covariant derivatives then reads\footnote{$m$ is in general complex but for our purposes, its imaginary part plays no role.}: 
\begin{align}
\{d_{\theta,\alpha},\bar{d}_{\vartheta,\dot{\beta}}\}&=\partial_{\alpha\dot{\beta}}\label{ddx}\\
\{d_{\theta,\alpha},d_{\vartheta,\beta}\}&=\bar{m}C_{\alpha\beta}\label{ddmb}\\
\{\bar{d}_{\theta,\dot{\alpha}},\bar{d}_{\vartheta,\dot{\beta}}\}&=-mC_{\dot{\alpha}\dot{\beta}}\label{ddm}\\
[d_{\vartheta,\alpha},d_y]&=-d_{\theta,\alpha}\label{dyd}\\
[\bar{d}_{\vartheta,\dot{\alpha}},d_y]&=-\bar{d}_{\theta,\dot{\alpha}}\label{bdyd}
\end{align}
Such an algebra can be incorporated in the superspace by introducing additional bosonic coordinates corresponding to the central charges. Then, requiring a trivial dependence of the hyperfields on these coordinates leads to a volume element same as the one when $m=0$. However, this generates explicit appearances of $\theta$'s in the Lagrangian (for example, last reference in \cite{GIKOS}).

There are two alternative manifestly covariant approaches to deal with non-zero $m$. One (simplest) approach is the dimensional reduction of 6D, N=1 massless multiplets to 4D, N=2 massive ones. Since projective superspace in 6D exists\cite{SJGJ} and is similar to the projective hyperspace in 4D, the main results can be written down just by inspection. We will show that this is the case in section \ref{secDR}, where we will compare the results derived via another approach.

In this second approach, we stay in 4D and turn $d$'s into covariant derivatives: $\D=d+A$, where $A$ is an Abelian connection that has acquired a vev, i.e. $A\propto m$. This avoids the explicit $\theta$'s in the Lagrangian that are now hidden inside the connections\cite{FGrey,IBSK}. So, the starting point for the simplest example of a massive hypermultiplet is a massless scalar hypermultiplet (SH) coupled to a U(1) vector hypermultiplet (VH).

Let us now briefly review the massless hypermultiplets living in projective hyperspace. The following discussion is valid in both 4D \& 6D with a few obvious changes, some of which will be pointed out later. A massless SH is represented by a complex `arctic' projective hyperfield ($\Upsilon$)\footnote{We use the arrow notation $\Psi\left[n^\uparrow_{(\downarrow)}\right]$ to denote that the hyperfield $\Psi$ contains $y^m$ with $m\geq(\leq)\,n$ only.}:
\be
d_\vartheta(\bar{d}_\vartheta)\Upsilon\left[0^\uparrow\right]=0 \Rightarrow  d_\vartheta(\bar{d}_\vartheta)\bar{\Upsilon}\left[1_\downarrow\right]=0.\label{pY}
\ee
Its on-shell expansion containing complex scalars $A\,\&\,B$ and Weyl spinors $\chi\,\&\,\tilde{\chi}$ is:
\be
\Upsilon=\(A+yB\)+\(\theta\chi+\bar{\theta}\bar{\tilde{\chi}}\)+\theta\partial B\bar{\theta}\label{osmsY}
\ee
and their corresponding equations of motion follow from $d_y^2\bar{\Upsilon}(\Upsilon)=0$, which in turn follow from the action:
\be
\S_{\Upsilon}=-\int dx d^4\theta dy \bar{\Upsilon}\Upsilon.
\ee

A VH is represented by a real `tropical' projective hyperfield ($V$):
\be
d_\vartheta(\bar{d}_\vartheta)V\left[0^\uparrow_\downarrow\right]=0.
\ee
Since we are mainly interested in the vevs of Abelian connections, we give below the vev structure of $V$ (read from the full expression for $V$ in Wess-Zumino gauge\cite{DJWS}) to which the connections will eventually get related:
\be
V=\frac{1}{y}\(\theta^2\bar{m}-\bar{\theta}^2m\).\label{Vvev}
\ee
Finally, the Lagrangian of a massless SH coupled to VH is simply given by:
\be
\S_{\Upsilon-V}=-\int dx d^4\theta dy \bar{\Upsilon}\exp{V}\Upsilon.\label{YbVY}
\ee

These are all the massless ingredients we need to construct the massive SH in projective hyperspace.

\newpage
\sect{4D Approach}

\subsect{Action\label{SHA}}
We have already argued that a massive SH is equivalent to a massless SH coupled to an Abelian VH with a vev. This means that we should be able to represent a massive SH by a complex projective hyperfield $\hat{\Upsilon}$. We start by writing a quadratic action for it that should be equivalent to eq. \ref{YbVY}:
\be
\S_{\hat{\Upsilon}}=-\int dx d^4\theta dy \bar{\hat{\Upsilon}}\hat{\Upsilon}=-\int dx d^4 \theta dy \bar{\Upsilon}\exp{V}\Upsilon.
\ee
The equations of motion for $\hat{\Upsilon}\,\&\,\Upsilon$ can be derived in a way similar to the massless case and they read:
\be
d_y^2\bar{\hat{\Upsilon}}=\D_y^2\bar{\Upsilon}=\int\frac{dy_2}{y_{12}}\bar{\Upsilon}_2\exp{V_2}=0.\label{eomYmv}
\ee

We know the massive equations of motion (Klein-Gordon \& Dirac equations) for the component fields and the expression for vev of $V$ (eq. \ref{Vvev}), so it is a simple algebraic exercise to get the (new) on-shell form of $\Upsilon$:
\be
\Upsilon=\(A+yB\)+\(\theta\chi+\bar{\theta}\bar{\tilde{\chi}}\)+\(\theta^2m-\bar{\theta}^2\bar{m}\)B+\theta\partial B\bar{\theta}.\label{osmvY}
\ee
This form (obviously) gives the correct massless limit (eq. \ref{osmsY}) when $m=0$. Plugging this expression in the action gives the usual kinetic terms for the component fields and the mass terms have an expected appearance:
\[\sim\int dx\(m\chi\chi+\bar{m}\bar{\tilde{\chi}}\bar{\tilde{\chi}}\)+m\bar{m}\(\bar{A}A+\bar{B}B\).\]
It is important to note that if we had na\"{\i}vely used the eq. \ref{osmsY} in above calculation, we would have gotten a wrong sign for $B$'s mass term! This small calculation makes it clear that we now have a correct representation for the massive SH. Thus, we can assign\footnote{$V=V_+[0^{\uparrow}]+V_-[0_{\downarrow}]$.} $\hat{\Upsilon}=\exp{V_+}\Upsilon$ and $\bar{\hat{\Upsilon}}=\bar{\Upsilon}\exp{V_-}$ such that their on-shell $y-$dependence remains the same as that of the massless hyperfields i.e. $\left[0^\uparrow\right]\,\&\,\left[1_\downarrow\right]$, respectively.

Moreover, in this case we can also figure out what $\D$'s look like explicitly. Comparing the two forms of equations in \ref{eomYmv} (with $\bar{\Upsilon}$), we get:
\begin{align}
\D_y^2\bar{\Upsilon}&=\partial_y^2\bar{\Upsilon}-\frac{2\(\theta^2\bar{m}-\bar{\theta}^2m\)}{y^2}\(\partial_y\bar{\Upsilon}-\frac{\bar{\Upsilon}}{y}\)-\frac{2\,\theta^2\bar{\theta}^2m\bar{m}}{y^4}\bar{\Upsilon}\\
\Rightarrow\D_y&=\partial_y+A_y={\partial}_y-\frac{\(\theta^2\bar{m}-\bar{\theta}^2m\)}{y^2}\\
\Rightarrow A_y&=d_y\int dy\'\frac{V\'}{(y-y\')}.
\end{align}
We can also find the expressions for other connections using equations \ref{dyd} \& \ref{bdyd} in the gauge $A_\vartheta=\bar{A}_\vartheta=0$:
\be
A_\theta=-d_{\vartheta}A_y=\frac{\bar{m}\theta}{y}\quad\&\quad\bar{A}_\theta=-\bar{d}_{\vartheta}A_y=\frac{-m\bar{\theta}}{y}.
\ee
These obviously satisfy the equations \ref{ddmb} \& \ref{ddm} as can be easily checked\footnote{For example, $d_\vartheta=\partial_\vartheta+y\partial_\theta+\bar{\theta}\partial_x$ and $\bar{d}_\vartheta=\bar{\partial}_\vartheta+y\bar{\partial}_\theta+\partial_x\theta$ in reflective representation.}. This completes the basic construction of a massive scalar hyperfield.

The coupling of this massive hypermultiplet to a non-Abelian VH\footnote{Having a central charge in the superalgebra does not make the vector hypermultiplet massive! This is because $\int d^2\theta d^2\vartheta\,W^2\xrightarrow{m\neq 0}\int d^2\theta d^2\vartheta(W+m)^2=\int d^2\theta d^2\vartheta\,W^2$. The equality holds because $\int d^2\theta d^2\vartheta\,W$ is a total spacetime derivative due to the Bianchi identity $d_\theta^2W=\bar{d}_\theta^2\bar{W}$.} is a straightforward generalization similar to the case of massless SH:
\be
\S_{\hat{\Upsilon}-\hat{V}}=-\int dx d^4\theta dy \bar{\hat{\Upsilon}}\exp{\hat{V}}\hat{\Upsilon}.\label{SYVY}
\ee

\vspace{10pt}
\subsect{Propagator}
The quantization of massive SH action is almost identical to that of the massless SH. First, we need to rewrite the massive scalar hyperfield in terms of a generic unconstrained hyperfield:
\begin{align*}
\hat{\Upsilon}(y_2)\left[0^{\uparrow}\right]&=d_{2\vartheta}^4\int dy_1\frac{1}{y_{12}}\Phi(y_1)\left[0^{\uparrow}_{\downarrow}\right]\\
\mathrm{and,}\quad\bar{\hat{\Upsilon}}(y_2)\left[1_{\downarrow}\right]&=d_{2\vartheta}^4d_{y_2}^2\int dy_1\frac{1}{y_{21}}\bar{\Phi}(y_1)\left[0^{\uparrow}_{\downarrow}\right].
\end{align*}
Then, we add source terms to the action and convert the $d^4\theta$ integral to $d^8\theta$ integral by rewriting $\hat{\Upsilon}$ using above relations:
\be
{\S}_{\hat{\Upsilon}-\hat{J}}=-\int dx\, d^8\theta\int dy_1\left[d_{y_1}^2\int dy_3\frac{\bar{\hat{\Upsilon}}_3}{y_{13}}d_{1\vartheta}^4\int dy_2\frac{\hat{\Upsilon}_2}{y_{21}}+\bar{\hat{J}}_1\int dy_2\frac{\hat{\Upsilon}_2}{y_{21}}+d_{y_1}^2\int dy_3\frac{\bar{\hat{\Upsilon}}_3}{y_{13}}\hat{J}\right]\label{actU}
\ee
where the sources $\hat{J}\,\&\,\bar{\hat{J}}$ are generic projective hyperfields. The equation of motion for $\hat{\Upsilon}$ with the source reads:
\be
\int dy_1\frac{d_{1\vartheta}^4 d_{y_1}^2\hat{\Upsilon}_1}{y_{13}}=-\int dy_1 d_{1\vartheta}^4 d_{y_1}^2\left(\frac{1}{y_{13}}\right)\hat{J}_1.\label{eomUwi}
\ee
The difference with respect to the massless case arises at this stage due to the presence of central charges in the superalgebra, which gives the following modified identity:
\[d_\vartheta^4d_y^2d_\vartheta^4=\(\square-2m\bar{m}\)d_\vartheta^4.\]
Using this identity in eq. \ref{eomUwi} leads us to the following equations:
\begin{align}
\(\square-2m\bar{m}\)\hat{\Upsilon}_3=&-d_{3\vartheta}^4\int dy_1 \frac{2\hat{J}_1}{y_{13}^3} \label{eomU}\\
\mathrm{Similarly,}\,\,\(\square-2m\bar{m}\)\bar{\hat{\Upsilon}}_2=&-d_{2\vartheta}^4\int dy_1 \frac{2\bar{\hat{J}}_1}{y_{21}^3}.\label{eomUb}
\end{align}
Plugging these equations back in action \ref{actU}, we get:
\be
{\S}_{\hat{\Upsilon}-\hat{J}}=\int dx\,d^8\theta\,dy_1\,dy_2\left[\bar{\hat{J}}_1\frac{1}{y_{21}^3}\frac{1}{\frac{1}{2}\(\square-2m\bar{m}\)}\hat{J}_2\right].
\label{SYJ}
\ee
This leads to the expected change in the massless propagator to give us the massive SH propagator:
\be
\bigl\langle\hat{\Upsilon}(1)\bar{\hat{\Upsilon}}(2)\bigr\rangle=-\frac{d_{1\vartheta}^4d_{2\vartheta}^4\delta^8(\theta_{12})}{y_{12}^3}\frac{\delta(x_{12})}{\frac{1}{2}\square-m\bar{m}}.
\label{pmvY}
\ee

\vspace{10pt}
\subsect{Vertices}
As in the massless case, there are no self-interacting renormalizable vertices for massive SH. The interactions appear purely with the coupling to a VH as seen in action \ref{SYVY}. That means the vertices look similar to the massless case:
\[\bar{\hat{\Upsilon}}^i \hat{V}^{j_1}...\hat{V}^{j_n}\hat{\Upsilon}^k\,\rightarrow\,\int d^4\theta\int dy\left({}_i\line(1,0){15}\line(0,1){15}^{j_1}\line(-1,0){8}\line(1,0){10}...\line(1,0){10}\line(0,1){15}^{j_n}\line(-1,0){9}\line(1,0){15}{}_k\right)\]
where, the group theory factor shown in parentheses is for adjoint representation.

\vspace{20pt}
\sect{6D Approach\label{secDR}}
We now explain the simpler method for obtaining a 4D massive scalar hypermultiplet: Dimensional Reduction of a 6D massless SH\cite{SJGJ}. First of all, we dimensionally reduce the bosonic coordinates from 6D $\(X^{M=0...5}\)$ to 4D $\(x^{\mu=0...3}\)$ by defining a complex coordinate:
\be
z(\bar{z})=\frac{1}{\sqrt{2}}\left[X^4+(-)\I X^5\right]\quad\Rightarrow\quad \partial(\bar{\partial})\equiv\partial_z(\partial_{\bar{z}})=\frac{1}{\sqrt{2}}\left[\partial_4-(+)\I\partial_5\right]
\ee
 and demanding that the corresponding momenta equal the 4D central charges:
\[p=-\I\partial=m\quad\&\quad\bar{p}=-\I\bar{\partial}=\bar{m}.\]
The 6D d'Alambertian then reduces to:
\be
\square_{\underline{6}}=\partial^M\partial_M=\partial^\mu\partial_\mu+2\partial\bar{\partial}=\square_{\underline{4}}-2m\bar{m}.
\label{B64}
\ee

Secondly, we reduce the fermionic coordinates in 6D, which are represented by Weyl spinors of SU*(4)  to 4D coordinates, which are represented by dotted \& undotted Weyl spinors of SL(2,C):
\be
\Theta^{\tilde{\alpha}}=\bpm \theta^{\alpha}\\ \bar{\theta}^{\dot{\alpha}}\epm
\ee
with similar relation holding true for $\vartheta$'s. The charge conjugation in 6D works as follows:
\be
\bar{\Theta}^{\tilde{\alpha}}\equiv {C^{\tilde{\alpha}}}_{\dot{\tilde{\beta}}}\bar{\Theta}^{\dot{\tilde{\beta}}}=\bpm \theta^{\alpha}\\ -\bar{\theta}^{\dot{\alpha}}\epm.
\ee
The 6D, N=(1,0) algebra of supercovariant derivatives is equivalent to the 4D, N=2 algebra in equations \ref{ddx}-\ref{ddm}, after the dimensional reduction. Furthermore, we can express a vector using just spinorial indices in 6D too:
\be
V_{\tilde{\alpha}\tilde{\beta}}=\frac{1}{2}\bpm \bar{v}\,C_{\alpha\beta} & v_{\alpha\dot{\beta}}\\ v_{\dot{\alpha}\beta} & v\,C_{\dot{\alpha}\dot{\beta}}\epm
\ee
where $v\(\bar{v}\)\sim -\I\left[V_4+(-)\I V_5\right]$.

We are now ready to deal with the 6D, N=1 massless hypermultiplets. Like 4D, SH is represented by a projective arctic hyperfield $\Upsilon_{\underline{6}}$. Using the (bi)spinor matrices defined above, we can reduce the $\Upsilon_{\underline{6}}$ (eq. \ref{osmsY}) to 4D `massive' SH:
\begin{align}
\Upsilon_{\underline{6}}&=(A+yB)+\Theta\Xi+\bar{\Theta}^{\tilde{\alpha}}\partial_{\tilde{\alpha}\tilde{\beta}}B\Theta^{\tilde{\beta}}\\
\Rightarrow\Upsilon_{\underline{4}}&=(A+yB)+\(\theta\chi+\bar{\theta}\bar{\tilde{\chi}}\)+\(\bar{\theta}^{\dot{\alpha}}\partial_{\alpha\dot{\alpha}}B \theta^{\alpha}+\theta^2mB-\bar{\theta}^2\bar{m}B\),\nonumber
\end{align}
which is the same as in eq. \ref{osmvY}. A VH in 6D is again represented by a projective tropical hyperfield $V _{\underline{6}}$ and its lowest $\Theta-$component (in Wess-Zumino gauge) looks like:
\be
V_{\underline{6}}=\frac{\bar{\Theta}^{\tilde{\alpha}}A_{\tilde{\alpha}\tilde{\beta}}\Theta^{\tilde{\beta}}}{y}\Rightarrow V_{\underline{4}}=\frac{1}{y}\(\bar{\theta}^{\dot{\alpha}}A_{\alpha\dot{\alpha}}\theta^{\alpha}+\theta^2\bar{\phi}-\bar{\theta}^2\phi\).
\ee
If the scalar field $\phi$ develops a vev, then the above equation is identical to \ref{Vvev}. Moreover, the action of $\Upsilon_{\underline{6}}$ coupled to $V_{\underline{6}}$ is given by eq. \ref{YbVY} so the 6D hyperfields' reduction to 4D reproduces the same massive SH action derived in section \ref{SHA}.  

Now the propagator for $\Upsilon_{\underline{6}}$ is similar to that of the massless SH in 4D and the reduction to massive case is straightforward owing to eq. \ref{B64}:
\begin{align}
\bigl\langle\Upsilon_{\underline{6}}(1)\bar{\Upsilon}_{\underline{6}}(2)\bigr\rangle&=-\frac{d_{1\vartheta}^4d_{2\vartheta}^4\delta^8(\Theta_{12})}{y_{12}^3}\frac{\delta(X_{12})}{\frac{1}{2}\square_{\underline{6}}}\\
\Rightarrow\bigl\langle\Upsilon_{\underline{4}}(1)\bar{\Upsilon}_{\underline{4}}(2)\bigr\rangle\equiv\bigl\langle\hat{\Upsilon}(1)\bar{\hat{\Upsilon}}(2)\bigr\rangle&=-\frac{d_{1\vartheta}^4d_{2\vartheta}^4\delta^8(\theta_{12})}{y_{12}^3}\frac{\delta(x_{12})}{\frac{1}{2}\square_{\underline{4}}-m\bar{m}}\,,\nonumber
\end{align}
which is equivalent to eq. \ref{pmvY} derived from the 4D perspective.

\vspace{20pt}
\sect{Feynman Rules}
These are almost the same as those given in reference \cite{DJWS}. The only difference is the following modified identity:
\be
d_{1\vartheta}^4d_{2\vartheta}^4d_{1\vartheta}^4=y_{12}^2\left[\(\frac{1}{2}\square-m\bar{m}\)+y_{21}\(\bar{d}_{2\theta}d_x d_{2\theta}+md_{2\theta}^2-\bar{m}\bar{d}_{2\theta}^2\)+y_{12}^2d_{2\theta}^4\right]d_{1\vartheta}^4
\ee
The non-renormalization theorem for massless scalar hypermultiplet holds for the massive case also for straightforward reasons.

\bfig[h]\bc
\includegraphics[width=16.226cm,height=2cm]{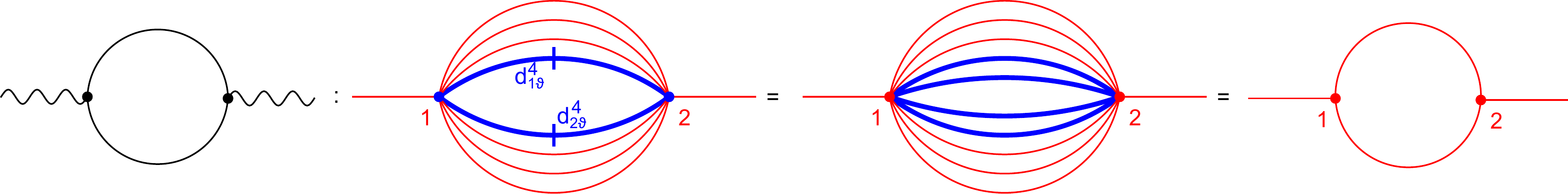}
\caption{One-hoop massive SH example with $d-$algebra \& $y-$calculus shown. [A blue (thick) line with a cut represents a $\delta^8({\theta}_{12})$.]}
\label{fig1}
\ec\efig

One-hoop correction to VH 2-point function (Figure \ref{fig1}) due to the coupling to a massive SH is simple to calculate and looks same (modulo the momentum integral) as the massless SH contribution:
\be
-\hat{\A}_2(p;m)\times c_R\,g^2\int d^8\theta\int dy_{1,2}\frac{\hat{V}_1\hat{V}_2}{y_{12}\,y_{21}}.\label{V1H2P}
\ee
The momentum integral is a standard integral and evaluates to (with $D=4-2\epsilon$):
\begin{align*}
\hat{\A}_2&=\int\frac{d^Dk}{(2\pi)^D}\frac{1}{\(\frac{1}{2}k^2-m\bar{m}\)\(\frac{1}{2}(k+p)^2-m\bar{m}\)}\\
&=\frac{1}{4\pi^2}\left[\frac{1}{\epsilon}-{\gamma}_{E}+2-\mathrm{ln}\(\frac{2m\bar{m}}{{\mu}^2}\)-\sqrt{1+\frac{8m\bar{m}}{p^2}}\,\,\mathrm{ln}\(\frac{1+p/\sqrt{p^2+8m\bar{m}}}{1-p/\sqrt{p^2+8m\bar{m}}}\)\right].
\end{align*}

\sect{Conclusion}
We presented a reformulation of the massive scalar hypermultiplet that allows derivation of the known results in a compact manner. Our analysis makes a massive scalar hypermultiplet more transparent at the component level. The diagrammatic Feynman rules are similar to the massless case and hence no extra effort is needed to evaluate diagrams with massive SH lines. We also presented an explicit 1-hoop calculation showing that the hypergraph rules allow computation as `fast' as the N=1 supergraph rules.

\newpage
\section*{\bc\color{sect}{Acknowledgements}\ec}
This work is supported in part by National Science Foundation Grant No. PHY-0969739.

\references{
\bibitem{ULMR}
U. Lindstr\"{o}m and M. Ro\v{c}ek, \cmp{115}{1988}{21};\\
U. Lindstr\"{o}m and M. Ro\v{c}ek, \cmp{128}{1990}{191}.
\bibitem{FGrey}
F. Gonzalez-Rey, M. Ro\v{c}ek, S. Wiles, U. Lindstr\"{o}m and R. von Unge, \npb{516}{1998}{426} [\arXivid{hep-th/9710250}];\\
F. Gonzalez-Rey and R. von Unge, \npb{516}{1998}{449} [\arXivid{hep-th/9711135}];\\
F. Gonzalez-Rey, 1997, \arXivid{hep-th/9712128}.
\bibitem{GIKOS}
A. Galperin, E. Ivanov, S. Kalitzin, V. Ogievetsky and E. Sokatchev, \cqg{1}{1984}{469};\\
E. Ivanov, A. Galperin, V. Ogievetsky and E. Sokatchev, \cqg{2}{1985}{601, 617};\\
A.S. Galperin, E.A. Ivanov, V.I. Ogievetsky, and E.S. Sokatchev, {\it Harmonic superspace} (Cambridge Univ. Press, 2001).
\bibitem{IBSK}
I. L. Buchbinder and S. M. Kuzenko, \cqg{14}{1997}{L157}.
\bibitem{WS-AdS}
W. Siegel, 2010, \arXivid{1005.2317}.
\bibitem{DJWS} 
D. Jain and W. Siegel, \pr{D}{80}{2009}{045024} [\arXivid{0903.3588}];\\
D. Jain and W. Siegel, \pr{D}{83}{2011}{105024} [\arXivid{1012.3758}].
\bibitem{SJGJ}
S. J. Gates, Jr., S. Penati and G. Tartaglino-Mazzucchelli, \jhep{05}{2006}{051} [\arXivid{hep-th/0508187}].}
\end{document}